\documentclass{aa}

\usepackage[varg]{txfonts}
\usepackage{graphicx}	
\usepackage{amsmath,amssymb}	
\usepackage{hyperref}
\usepackage{stfloats,placeins}
\newcommand{\kms}{\,km\,s$^{-1}$} 

\begin{document}

\title{The variation of intensity ratio at each wavelength point of the \ion{Si}{IV} 1394/1403 \AA\ lines.}
\subtitle{Spectral diagnostic of a bifurcated eruption}

\author{
	Yi'an Zhou\inst{1,2}
	\and Xiaoli Yan\inst{1,2}
	\and Zhike Xue\inst{1,2}
	\and Liheng Yang\inst{1,2}
	\and Jincheng Wang\inst{1,2}
	\and Zhe Xu\inst{1,2}
}
\institute{
	Yunnan Observatories, Chinese Academy of Sciences, Kunming 650216, China \\
	\email{zhouyian@ynao.ac.cn}
\and
	Yunnan Key Laboratory of the Solar physics and Space Science, Kunming 650216}

\date{Received 5 February 2024 / Accepted 25 October 2024}

\abstract{}
{This study aims to investigate the deviation of the intensity ratio of the \ion{Si}{IV} 1394 \AA\ and 1403 \AA\ emission lines from the expected value of 2 in the optically thin regime, as observed in many recent studies.}
{We analyzed the integrated intensity ratio ($R$) and the wavelength-dependent ratio ($r(\Delta\lambda)$) in a small bifurcated eruption event observed by the Interface Region Imaging Spectrograph (IRIS).}
{Despite the relatively complex line profiles, {most of the intensity ratio $R$ of \ion{Si}{IV} lines remained greater than 2 in the loops. The ratio $r(\Delta\lambda)$ varied in the line core and wings, changing distinctly from 2.0 to 3.3 along the wavelength.} At certain positions, the \ion{Si}{IV} 1394 \AA\ and 1403 \AA\ lines exhibited different Doppler velocities.}
{When diagnosing the spectra of small active region events, not only the impact of opacity but also the influence of resonance scattering should be considered. We propose that the ratio $r(\Delta\lambda)$ can serve as an indicator of the resonance scattering and opacity effect of the \ion{Si}{IV} line.}

\keywords{Sun: activity -- Sun: transition region -- Sun: UV radiation -- line: profiles -- opacity}

\maketitle

\section{Introduction}
In the solar transition region (TR), a multitude of emission lines are formed. These include the \ion{Si}{IV} 1394 \AA\ and 1403 \AA\ lines, the \ion{O}{IV} 1400 \AA\ and 1401 \AA\ lines, and the \ion{C}{IV} 1548 \AA\ and 1550 \AA\ lines. These emission lines play a pivotal role in unveiling the physical properties of the TR during various microscale events. These events include transient brightenings \citep{Muller2003,Bahau2021,Chitta2021}, explosive events \citep{Teriaca2004,Huang2014,Gupta2015,Huang2017,Chen2019a}, nanoflares \citep{Testa2014,Tian2014}, and Ellerman Bombs \citep{Vissers2015,Tian2016,Chen2019b}.

The ratios of these TR emission lines offer a multitude of advantages. For instance, the \ion{O}{IV} line ratio is particularly beneficial for density diagnostics \citep{Doschek1984, Cook1995}. Furthermore, some emission lines (e.g. \ion{Si}{IV} 1394/1403 \AA\ and \ion{C}{IV} 1548/1550 \AA\ lines) are originated from the same plasma, these line pairs are ought to have similar line profiles. Thus the \ion{Si}{IV} line ratio is effective for analyzing opacity effects. Given the low density of the upper atmosphere in the quiet sun, it is generally assumed that the emitted \ion{Si}{IV} photons can escape without further scattering and absorption. Under these circumstances, the ratio of \ion{Si}{IV} resonance lines should be precisely 2 in the optically thin regime \citep{Maniak1993,Math1999}, which equals to {the ratio of} their  oscillator strengths.     

During various solar activities, the \ion{Si}{IV} lines tend to exhibit a significant enhancement compared with those in the quiet sun, and their intensity ratio may deviate from 2. The \ion{Si}{IV} lines may be subject to a self-absorption effect during transient brightenings, leading to a decrease in the intensity ratio from 2 to approximately 1.7 \citep{Yan2015,Nelson2017}. A comprehensive statistical study by \cite{Tripathi2020} further indicates that the intensity ratios of the \ion{Si}{IV} lines within an emerging flux region display significant temporal and spatial variations. In the early stages of the region’s evolution, the intensity ratios are predominantly less than 2. {\cite{Babu2024} also proposed that the intensity ratio deviates from 2 more than half of the footpoints of the cool loops.}
As the active region (AR) evolves, a multitude of ratios approach or exceed 2. Moreover, resonance scattering can also result in a ratio of the \ion{Si}{IV} resonance lines greater than 2 \citep{Gontokakis2018}.
In addition, \cite{Peter2014} reported that the ratio of \ion{Si}{IV} lines is near 2 in a hot explosive event, but there is a distinct dip in the centroid of the \ion{Si}{IV} 1394 \AA\ line. Furthermore, \cite{Kerr2019} conducted a detailed numerical simulation and confirmed that the \ion{Si}{IV} lines could be optically thick in flaring scenarios. The simulation demonstrated that the intensity ratio of \ion{Si}{IV} lines may range from 1.8 to 2.3 at flare ribbons, which aligns with many recent observations \citep{Brannon2015,Mulay2021,Wang2023}.

Despite the prevalent use of the intensity ratio as a primary criterion for evaluating opacity effects, its effectiveness remains a topic of ongoing discussion. This is due to the fact that the core or wings of the \ion{Si}{IV} lines may retain their optically thin nature even when the integrated ratio diverges from 2. That is to say, some photons of the \ion{Si}{IV} lines can still escape from the core or wings even though the intensity ratio is greater or less than 2. \cite{Zhou2022} demonstrated that the ratio of intensity at each wavelength point serves as a valuable diagnostic tool for assessing the opacity effect. They observed that the \ion{Si}{IV} line profiles exhibited a central reversal, characterized by a depth equivalent to half of the peak intensity, at numerous positions during an M7.3 class flare. The integrated ratio, denoted as $R$, is close to 2. However, the ratio along the wavelength, represented as $r(\Delta\lambda)$, displayed variability from the line wings to the core. The ratio $r(\Delta\lambda)$ is low to around 1.3 near the line core and displayed obvious line opacity. This observation aligns with the result of \ion{Si}{IV} lines in radiative hydrodynamics flare models that the ratio $r(\Delta\lambda)$ profile performs a U shape, in which $r(\Delta\lambda)$ decreases from both blue and red line wings to the line core \citep{Yu2023}. Nevertheless, for events beyond flares, such as TR explosive events, minor jets, or small eruptions, further research is warranted to elucidate the relationship between the ratios $R$ and $r(\Delta\lambda)$ of \ion{Si}{IV} lines and opacity.

In this paper, we investigate the two ratios ($R$ and $r(\Delta\lambda)$) of \ion{Si}{IV} lines in a small bifurcated eruption event observed by the Interface Region Imaging Spectrograph (IRIS; \cite{Depont2014}). The observations by IRIS are conducted in a scanning mode. Furthermore, due to the relatively complex structure of this eruption, we also utilized images from the Atmospheric Imaging Assembly (AIA;\cite{Lemen2012}) and magnetograms from the Helioseismic and Magnetic Imager (HMI;\cite{Scherrer2012}) of Solar Dynamics Observatory (SDO; \cite{Pesnell2012}) to aid in understanding the evolution of the eruption. The structure of our paper is as follows. In Section \ref{sec-obs-data}, we primarily introduce the overview of the small eruption and the data reduction. Section \ref{res} provides a detailed analysis of the \ion{Si}{IV} line profiles and the two associated ratios ($R$ and $r(\Delta\lambda)$). In the Sections \ref{sec_discuss} and \ref{sec_conclu}, we discuss and summarize the results, respectively.

\section{Observations and data reduction}
\label{sec-obs-data}

\begin{figure*}
	\centering
	\includegraphics[width=\hsize]{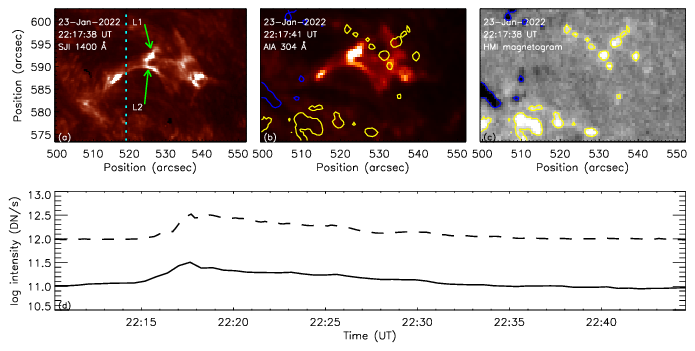}
	\caption{Panels (a)-(c) : Overview of the bifurcated eruption around 22:17:40 UT in NOAA AR 12932 acquired at SJIs 1400 \AA\ , AIA 304 \AA\ and HMI magnetogram, respectively. In panel (a), isolated loops are marked by L$_1$ and L$_2$. Cyan dotted line in SJIs indicate the slit position at 22:17:38 UT. The yellow and blue isogauss contours in panels (b) and (c) are the levels of $\pm$50 G, respectively. Panel (d) : The time profiles of the total intensity of the AIA 304 \AA\ (dashed) and SJI 1400 \AA\ (solid) during the eruption.}
	\label{fig1}
\end{figure*}

In this paper, we focus on the small bifurcated eruption occurred in NOAA active region (AR) 12932. Started at 22:15 UT, the eruption with low intensity lasted for about 15 mins and faded away near 22:30 UT on 2022 Jan 23. IRIS provides a high-cadence and large dense 320-step raster observation of this eruption, with a slit width of 0\farcs 33 and step cadence of 9.2 s. The time cadence of each 1400 \AA\ and 2796 \AA\ slit-jaw images (SJIs) is $\sim$36.7 s, and the time cadence of spectra is $\sim$9.0 s. We mainly use the \ion{Si}{IV} resonance lines in the far-ultraviolet (FUV) waveband of IRIS spectra for study. The spectral resolution of this waveband is $\sim$0.025\AA. We also illustrate the 304 \AA\ images for this active region, with a time cadence of 12 s and spatial resolution of 1\farcs 2. Besides, the line-of-sight magnetic field of HMI is also used for study. Figure \ref{fig1}{(a)-(c)} illustrates an overview of the bifurcated eruption with SJI 1400\AA\ , AIA 304 \AA\ and HMI magnetograms at around 22:17:40 UT, and Figure \ref{fig1}{(d)} shows the time profiles of the total intensity of the eruption.

In the initial stage of the observed activity, distinct brightenings were discernible in the field of view of SJI 1400 \AA\ at $\sim$ 22:15 UT. Following this, a loop-like structure {L$_0$} emerged in the same area $\sim$ 22:16 UT. This structure subsequently bifurcated into two components, denoted as L$_1$ and L$_2$, which are illustrated in Figure \ref{fig1}(a) and (b). These components reached their peak intensity at $\sim$ 22:17 UT. The line-of-sight magnetic field of HMI, as depicted in Figure \ref{fig1}(c), clearly demonstrates that the footpoints of each loop are proximal to the positive/negative poles. This is further corroborated by the overplotted HMI contours in Figure \ref{fig1}(b). The northern loop (L$_1$) began to dissipate $\sim$ 22:20 UT, while the southern loop (L$_2$) persisted for a slightly longer duration before eventually disappearing at $\sim$ 22:27 UT. 

The IRIS slit traversed the loop structures at various instances. We predominantly select the IRIS slit data around $\sim$ 22:17 UT for analysis. When using the line profile, we chose the average of the data within $\pm$1 pixel at the slit as the reference profile. The reference centroid of the \ion{Si}{IV} resonance lines was calibrated using the photospheric S I 1401.5 \AA\ line.
We also calculate the Doppler velocities of \ion{Si}{IV} 1394 and 1403 \AA\ lines with the moment method, {and their corresponding errors are calculated.}
We then employed multi-Gaussian profiles to fit the components of the two \ion{Si}{IV} 1393.7 and 1402.8 \AA\ resonance lines on the loops. {The fitted velocities are then derived from the individual components of the Gaussian profiles, and the errors are determined from the perror parameter in the mpfitfun function.} 
Furthermore, we calculate the integrated ratio ($R=\int \mathrm{I}_{1394}(\lambda)\mathrm{d}\lambda/\int \mathrm{I}_{1403}(\lambda)\mathrm{d}\lambda$) and the ratio of intensity along the wavelength point ($r(\Delta\lambda)=\mathrm{I}_{1394}(\Delta\lambda)/\mathrm{I}_{1403}(\Delta\lambda)$) of the \ion{Si}{IV}  resonance lines to investigate the opacity effect of the loops. 
To enhance the precision of our calculations, we excluded the saturated data and {blended lines}.
{Additionally, we subtracted the continuum (\text{cont(time)}) at each special moment within the \ion{Si}{IV} resonance lines. The value of \text{cont(time)} is equal to the mean intensity of \ion{Si}{IV} 1394 \AA\ and 1403 \AA\ lines at ten relatively quiet positions in the slit at each moment.} 
The selected range for the ratio of \ion{Si}{IV} lines was the wavelength where the intensity exceeded thrice the standard deviation of the far line wing intensity fluctuation. 

Moreover, the effective area of IRIS exhibits wavelength dependence\citep{Wulser2018}. 
{We utilized the latest version of Solarsoft routine iris\_getwindata.pro \citep{Young2015} to convert the obtained IRIS spectra from units of DN s$^{-1}$ to physical units of erg\ cm$^{-2}$\ s$^{-1}$\ sr$^{-1}$ to get the radiometric calibrated data, as well as the intensity uncertainties.
}

\section{Results}
\label{res}
\begin{figure*}
	\centering
	\includegraphics[width=2\columnwidth]{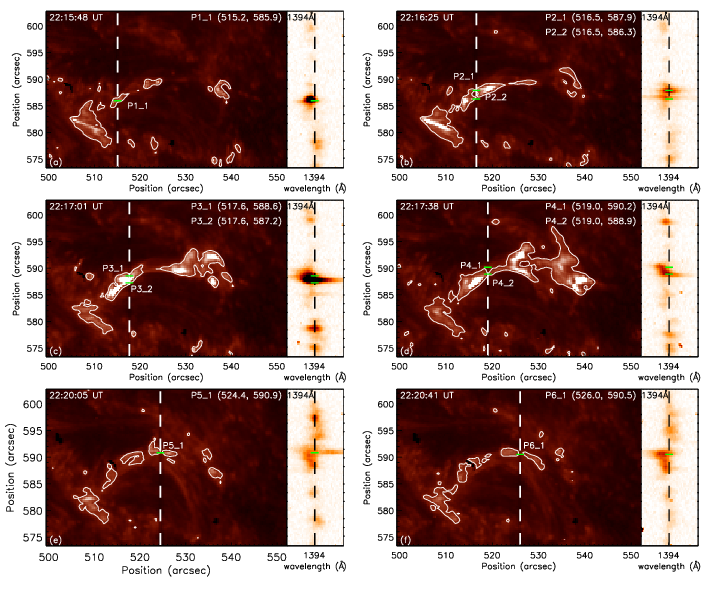}
	\caption{The IRIS SJI 1400 \AA\ images and the 1394 \AA\ spectra during the eruption. Within SJI 1400 \AA\ , the positions denoted by Px\_y and highlighted with green short lines are selected for further study. The white contours represent areas with an intensity thrice the average of the quiet region.  The white dashed lines in SJI images denote the position of slit. Black dashed lines in the spectra indicate the line center of \ion{Si}{IV} 1394 and 1403 \AA\ , while the short green lines indicate the position of Px\_y, respectively. {The precise coordinates of each Px\_y are shown in each panel.}}
	\label{fig2}
\end{figure*}

\begin{figure*}
	\centering
	\includegraphics[width=2\columnwidth]{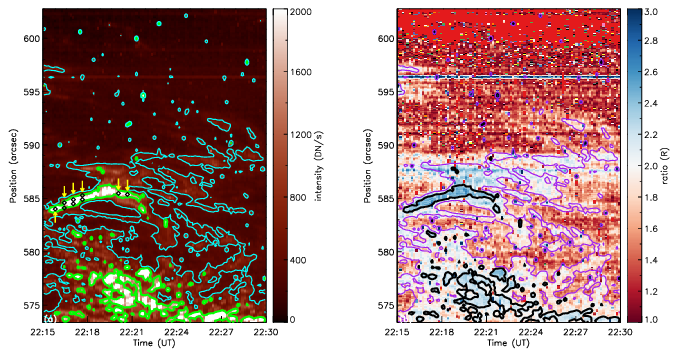}
	\caption{Panels (a) and (b) : The space-time maps of the total intensity of the \ion{Si}{IV} 1394 \AA\ line and the integrated ratio $R$, respectively. {The green/black and cyan/purple contours in each panel indicate areas with an intensity thrice the average of the quiet region and 0.05 times the maximum intensity of observed region, respectively.} The selected positions for study are marked by black diamonds and yellow arrows.}
	\label{fig3}
\end{figure*}

In this section, we choose the positions with distinct features on the spectra of the \ion{Si}{IV} resonance lines when the slit was moving across the loop-like structures (Figure \ref{fig2}, {labeled Px\_y})), and marked them with short green lines. 
{Here, x refers to the number of the slit selected, and y refers to the position studied within that slit. Additionally, we have indicated the precise coordinates of each Px\_y in Figure \ref{fig2}.}
Subsequently, we undertake a comprehensive examination of the spectral characteristics associated with the \ion{Si}{IV} lines at these designated positions. The green and black contours in Figure \ref{fig3} indicate the temporal variation of the line intensity of \ion{Si}{IV} 1394 \AA\ line in loop-like structures scanned by the IRIS slit, while the selected positions in Figure \ref{fig2} are marked with black diamonds and yellow arrows. 
The space-time map of the integrated ratio $R$ (Figure \ref{fig3}{(b)}) shows the ratio $R$ is mostly greater than 2 on the loops. 
{Additionally, the cyan and purple contours in Figure \ref{fig3} delineate regions within the field of view where the intensity exceeds 0.05 times the maximum intensity. It is apparent that, within the relatively quiescent regions of the field, the integral ratio $R$ predominantly hovers around 2, which is consistent with the expectations for an optically thin medium. In contrast, regions outside the cyan and purple contours exhibit weak intensities, leading to significant uncertainties and diminished reliability in the derived integral ratio $R $.}

\subsection{Spectral features at loops prior to eruption}
\label{sec31}
As shown in the spectra of Figure \ref{fig2}{(a) and (b)}, the selected positions {(P1\_1 at L$_0$, P2\_1 at L$_1$ and P2\_2 at L$_0$, respectively)} show obvious enhancement at 22:15:48 UT and 22:16:25 UT. 

\begin{figure*}
	\centering
	\includegraphics[width=2\columnwidth]{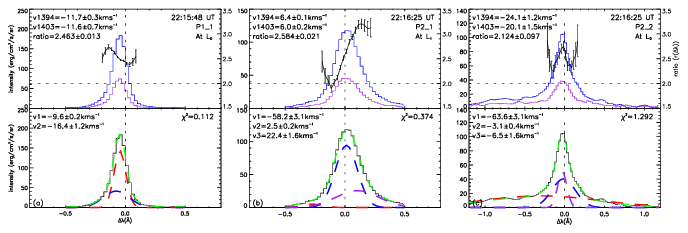}
	\caption{Line profiles of \ion{Si}{IV} 1394 \AA\ (blue) and 1403 \AA\ (purple) at loop positions prior to the bifurcated eruption. In upper panels, the Doppler velocity and integrated ratio $R$ are derived from moment analysis. The ratio $r(\Delta\lambda)$ are plotted with error bars. The vertical dashed lines indicate the line center of two \ion{Si}{IV} resonance lines, while the dashed horizontal lines are for $r(\Delta\lambda)=2$. The lower panels show the observed \ion{Si}{IV} 1394 \AA\ line (histogram) and fitted line profiles (red, blue and purple) at each positions, the velocities $\mathrm{v}_{1}$,$\mathrm{v}_{2}$ and $\mathrm{v}_{3}$ are derived from Gaussian fitting of each components. The green dashed lines indicate the combined Gaussian fitting. The $\mathrm{\chi}^{2}$ values in each panel have been divided by DOF (number of degrees of freedom).}
	\label{fig5}
\end{figure*}

The spectra in Figure \ref{fig2}{(a)} displays that the \ion{Si}{IV} line is mostly enhanced at the blue wing at P1\_1. 
We plot the profiles of \ion{Si}{IV} 1394 and 1403 \AA\ lines at P1\_1 in the upper panel of Figure \ref{fig5}{(a)} in blue and purple, respectively. 
The $r(\Delta\lambda)$ then added with error bars, the error of $r(\Delta\lambda)$ is  $\sqrt{(\delta \mathrm{dI}_{1394}/\mathrm{dI}_{1394})^{2}+(\delta \mathrm{dI}_{1403}/\mathrm{dI}_{1403})^{2}}$, in which the $\delta \mathrm{dI}$ is equal to the standard deviation of the intensity fluctuations in the far wings of \ion{Si}{IV} 1394 and 1403 \AA\ lines. In this study, the uncertainties owing to the IRIS instrument are significantly smaller than the observed uncertainties.
For the profiles at P1\_1, it is clearly seen that both the \ion{Si}{IV} 1394 and 1403 \AA\ lines are blueshifted with velocities of $\sim$ 12 \kms . 
By applying a double Gaussian fit to the \ion{Si}{IV} 1394 \AA\ line (lower panel of Figure \ref{fig5}{(a)}), 
{we see two distinct components with velocities of $\sim$ 10 and 16 \kms\ on the blue side. The ratio at each wavelength point ($r(\Delta\lambda)$) varies from the blue wing to the red wing, and shows a tendency to increase first and then decrease. At this position, the integrated ratio $R$ is around 2.5. }

In reference to position P2\_1 {at L$_1$}, as depicted in Figure \ref{fig2}{(b)}, there is a discernible enhancement at both the rest position and the {red/blue} wings of the \ion{Si}{IV} lines. This observation substantiates the assertion that the \ion{Si}{IV} resonance lines exhibit an overall redshift (Figure \ref{fig5}{(b)}). Furthermore, the velocities associated with these lines are approximately within the range of several \kms . Upon applying a {triple} Gaussian fit to the \ion{Si}{IV} 1394 \AA\ line, it becomes evident that the profile at this location can be bifurcated into {three} distinct components: a quasi-stationary component, a redshift {and relatively weak blueshift feature}. {It is also noteworthy that while the ratio $R$ remains larger than 2, but contrary to the P1\_1, the ratio $r(\Delta\lambda)$ decreases initially and then increases along the wavelength.}

The spectra at P2\_2 {at L$_0$} presents a distinct contrast to that at P2\_1 (Figure \ref{fig2}{(b)}). At this location, the profiles of the \ion{Si}{IV} line exhibits a conspicuous enhancement in the blue wing. {While the ratio $R$ is 2.1, the variation of the ratio $r(\Delta\lambda)$ displays a W shape, with its peak near the line center.} Furthermore, the spectral linewidth is markedly broader than those observed at P1\_1 and P2\_1. Upon employing a triple Gaussian fit to the \ion{Si}{IV} 1394 \AA\ line, it is discerned that the line possesses both a blueshift component, approximately 64 \kms, and two quasi-stationary components.

\subsection{Spectral features at loops during the eruption}
\label{sec32}
The spectra of \ion{Si}{IV} lines in Figure \ref{fig3}{(c)} illustrates that the radiation from the loop-like structure is significantly enhanced in both the red and blue wings at 22:17:01 UT. By examining the line profiles in Figure \ref{fig6}{(a) and (b)}, it can be further discovered that although the loop has not yet {fully} bifurcated at this time, there are distinct differences in the spectral characteristics of the lines at different positions (deviation of $\sim$ 1\farcs 2) on the loop. The \ion{Si}{IV} line profiles are blue-shifted at P3\_1 {at L$_1$}, while it is slightly red-shifted at P3\_2 {at L$_2$}. After applying \ion{Si}{IV} 1394 \AA\ line profiles at these two positions with a triple Gaussian fit, one can see that several components at position P3\_1 had blue-shifted velocities (Figure \ref{fig6}{(a)}), while there are {two obvious red-shifted and a blue-shifted velocity components at position P3\_2 (Figure \ref{fig6}{(b)}).} 
{
From the fitted line profile in Figure \ref{fig6}{(a) and (b)}, one can see that positions P3\_1 and P3\_2 share a similar velocity component (indicated by the purple dashed lines). The other velocity components (represented by the red/blue dashed lines) originate from bifurcated loops L$_1$ and L$_2$.}

{Unlike the ratio $r(\Delta\lambda)$ discussed in the section \ref{sec31}, the ratio $r(\Delta\lambda)$ at P3\_1 gradually increases from the blue wing to the red wing. The variation of $r(\Delta\lambda)$ near the line center at P3\_2 is the similar to that at P3\_1, with $r(\Delta\lambda)$ being basically near 2.5, but decreasing rapidly on the far blue wing. Nevertheless, the integrated ratio $R$ at these two positions is still near 2.5, similar to that at P1\_1 and P2\_1.}

At 22:17:38 UT (Figure \ref{fig3}(d)), the loop has already bifurcated. Meanwhile, the \ion{Si}{IV} line at position P4\_1 {at L$_1$} is significantly enhanced at the blue wing, while the enhancement of that at position P4\_2 {at L$_2$} is mostly at the red wing. By examining the \ion{Si}{IV} line profiles (Figure \ref{fig6}{(c) and (d)}), it can be easily discovered that the \ion{Si}{IV} 1394 \AA\ line at position P4\_1 is blue-shifted as a whole, with a velocity of about 23 \kms. On the other hand, the \ion{Si}{IV} line at position P4\_2 could be clearly split into two components with red and blue shifts, respectively, with Doppler velocities around 10 \kms . 
{Similar with the features at P3\_1 and P3\_2, the characteristic differences of the \ion{Si}{IV} lines at P4\_1 and P4\_2 are owing to the bifurcated loops L$_1$ and L$_2$.}
Another noteworthy fact is that there is a Doppler velocity difference of about 10 \kms between the \ion{Si}{IV} 1394 \AA\ and 1403 \AA\ lines at these two positions.

At the position P4\_1, the ratio $R$ is still greater than 2, while the ratio $r(\Delta\lambda)$ exhibits a noticeable variation from the blue wing to the red wing. 
{The ratio $r(\Delta\lambda)$ is significantly decreasing from the far red wing to the blue wing.
Regarding position (P4\_2), although the ratio $R$ remains greater than 2, the trend of $r(\Delta\lambda)$ is evident: it initially increases from 2.0 to 2.8 and subsequently decreases back to 2.0.}

\subsection{Spectral features at loops in the later stage of the eruption}

At 22:20:05 UT, the loop-like structure has completely fragmented. 
The IRIS slit successfully captured the position P5\_1 near the foot point of loop L$_1$, as depicted in Figure \ref{fig3}(e). 
The spectra reveals a substantial width of \ion{Si}{IV} line profile, with notable enhancements of the line emission discernible at both the red and blue wings, especially at the red wing. 
In correlation with the \ion{Si}{IV} line profiles shown in Figure \ref{fig7}{(a)}, one can see a prominent peak at the red wing, exhibiting a velocity $\sim$ 52 \kms. 
As shown in Figure \ref{fig7}{(b)}, the \ion{Si}{IV} line profiles at P6\_1 bear a resemblance to the blueshifted part of the profiles observed at P5\_1. 
{Considering that the location of P5\_1 is close to P6\_1 (Figure \ref{fig3}(f)), it is plausible that the redshift component at P5\_1 (blue dashed line in Figure \ref{fig7}{(a)}) persists for a duration spanning tens of seconds before eventually dissipating. }
Furthermore, there is a velocity difference of approximately 10 \kms between two \ion{Si}{IV} lines at these two positions, which is strikingly similar to the situation at P4\_2.

Despite the complexity of the \ion{Si}{IV} line profile at P5\_1, the integrated intensity ratio $R$ remains {greater than 2, at near 2.5, which is similar to the situation described earlier}. The ratio $r(\Delta\lambda)$ exhibits significant fluctuations along the wavelength points. 
At P6\_1, the line profile of \ion{Si}{IV} is relatively regular, with the ratio $r(\Delta\lambda)$ demonstrating {a similar tendency to that at P4\_2.}

\section{Discussion}
\label{sec_discuss}
{Intriguingly, we observed that most of the integrated intensity ratio $R$ exceeds 2 in the loop-like structures (Figure \ref{fig2}{(b)}).
This feature can be attributed to the unique geometric structures that contribute to opacity, as proposed by \cite{Rose2008}. Such structures can cause deviations in the integrated intensity ratio of resonance lines from their values in an optically thin environment.
We further estimated the electron density using the ratio of \ion{O}{IV} 1399.8 and 1401.2 \AA\ lines \citep{Dere1997, Dudik2014}. For the relatively weak \ion{O}{IV} lines, we bined 10 areas, each measuring 8$\times$4 (pixels$\times$pixels). In 8 of these regions,
the ratio $R$ exceed 2, while in the remaining 2 regions, it was less than 2. 
The electron density in the regions where the ratio $R$ was greater than 2 ranged around $10^{9.6}$ to $10^{10.2}$ cm$^{-3}$. Conversely,  in regions where the $R$ was less than 2, the electron density ranged from $10^{11.3}$ to $10^{11.7}$ cm$^{-3}$ . These results ailgn with those of \cite{Gontokakis2018}. 
Their simulation indicates that when the collisional excitation of electrons gradually decreases and resonance scattering dominates, the integrated intensity ratio of \ion{Si}{IV} lines will be greater than 2. 
Our observation is also consistent with cases where $R$ exceeds 2, such as in emerging flux regions \citep{Tripathi2020} and at the footpoints of cool loops \citep{Babu2024}.
The detailed physical mechanisms leading to a ratio greater than 2 still require further radiative hydrodynamic simulations and observations.}

Typically, the two \ion{Si}{IV} resonance lines, due to their proximate formation heights, should exhibit highly similar line profiles. 
{This similarity should ideally extend to the Doppler velocities derived from fitting these two lines, resulting in close values (like P1\_1 and P2\_1). However, an intriguing deviation from this expectation is observed at some positions during small eruptions, where the Doppler velocities of the two \ion{Si}{IV} lines often exhibit differences ranging from several to tens of \kms (from P2\_2 to P6\_1).
This feature is not unique to the \ion{Si}{IV} lines. Previous observations have identified a similar feature for the \ion{C}{IV} 1548/1550 Å lines in the active region, which also originate in the transition region \citep{Gontokakis2013, Gontokakis2016}. The shape differences between these resonance lines suggest that their line profiles are not solely generated by collisional excitation.
Upon synthesizing the above discussions, it is inferred that these differences primarily arise from resonance scattering. Furthermore, in certain off-limb events, similar resonance lines may also be influenced by linearly polarized light, leading to differences in their profile shapes \citep{Raouafi1999, Tavabi2015}. In addition, a comparison of the oscillator strength ratio of the \ion{Si}{IV} line with their collisional excitation strength, followed by a subsequent comparison with the situation of the \ion{C}{IV} lines \citep{Landi2012}, leads us to conjecture that the asymmetric structure of the \ion{Si}{IV} lines (like P4\_2 and P6\_1) may also be a result of resonance scattering.
}

{In recent studies, both observational and simulation data have demonstrated that the wavelength-dependent ratio, denoted as $r(\Delta\lambda)$, serves as an effective indicator of line profile opacity across various wavelengths. This ratio has been found to provide a more accurate representation of line opacity compared to the integrated ratio $R$ \citep{Zhou2022, Yu2023}.
In this work, we observed a notable deviation in the distribution of the ratio $r(\Delta\lambda)$ when compared to that observed in flare ribbons. Specifically, at several positions characterized by multiple velocity components, the $r(\Delta\lambda)$ profile exhibited vairous of shapes, with values fluctuating between 2.0 and 3.3. This observation suggests significant differences in opacity at each wavelength during this eruption. \cite{Yu2023} proposed that under flare conditions, the ratio $r(\Delta\lambda)$ exhibits a negative correlation with optical depth, while it shows a positive correlation with the resonance scattering effect. Given that the $r(\Delta\lambda)$ in our observations is predominantly greater than 2, we infer that resonance scattering is the dominant factor in this small eruption, leading to an overall increase in the ratio $r(\Delta\lambda)$.
Interestingly, our observations (as depicted in Figure \ref{fig6}{(d)} and \ref{fig7}{(b)}) reveal that the $r(\Delta\lambda)$ ratio in the bifurcated loop L${_2}$ (at positions P4\_2 and P6\_1) is smaller than that at the line center. This suggests that the opacity effects in the line wings exert a greater impact than those at the line core, which presents a contrast to the conditions typically observed in flares. }

Another interesting observation is the behavior of the ratio $r(\Delta\lambda)$ at positions where $R$ approaches 2 (Figure \ref{fig5}{(c)}). 
{In these instances, $r(\Delta\lambda)$  exhibits noticeable variations, yet it remains close to 2 at certain line wing positions.} This holds true even in the presence of a discernible multi-peak structure within the \ion{Si}{IV} lines at certain locations. 
This features present a contrast to the conditions observed within flare ribbons, yet bears similarity to those within flare loops. It suggests that, despite varying opacities across the \ion{Si}{IV} lines, photons originating from other optically thick line wings or core have the ability to scatter towards the {line wing's} vicinity and subsequently escape. 
In essence, these \ion{Si}{IV} lines can be considered quasi-optically thin to a certain degree, given that the majority of the photons within these resonance lines are neither absorbed nor re-emitted.

A noteworthy observation pertains to the discernible correlation between certain red- and blue-shifted components within the \ion{Si}{IV} line and the bifurcated loops. Specifically, the northern loop, denoted as L${_1}$, aligns with the blueshifts, whereas the southern loop, referred to as L${_2}$, corresponds to the redshifts observed in the \ion{Si}{IV} line. Interestingly, these red- and blue-shifted components occasionally coexist at identical positions on the loop. Their Doppler velocities, which can reach up to tens of \kms , may signify the presence of bidirectional flows within the loops when viewed in the line-of-sight \citep{Zhou2017,Polito2020}.
{However, from P3\_1 at L${_1}$ and P3\_2 at L${_2}$, as well as P4\_2 at L${_2}$ and P6\_1 at L${_1}$ (as dipicted in  Figure \ref{fig6}{(d)} and Figure \ref{fig7}{(b)}), it can be observed that the shape and variation of the ratio  $r(\Delta\lambda)$ are quite similar within a range of approximately $\pm$0.3 \AA\  around the line center. However, these two sets of positions exhibit significant differences in red- and blue-shifted components. Specifically, P3\_1 shows an overall blueshift, while P3\_2 has a noticeable redshift component. In contrast, P4\_2 exhibits a significant redshift component, while P6\_1 shows a clear blueshift component. This suggests that the distribution of the ratio $r(\Delta\lambda)$ is not directly correlated with these velocity components. Nevertheless, at far red and blue wing measurements, where the ratio $r(\Delta\lambda)$ is smaller and calculation errors are larger, we still require additional observational evidence to discuss whether there exists a relationship between  $r(\Delta\lambda)$ and Doppler velocities.}
At certain positions, the line profiles of \ion{Si}{IV} are also characterized by a high-intensity, narrow-width core, accompanied by a weaker, albeit wider, wing. This particular structure of spectral lines is frequently observed in emission lines originating from the middle TR, which has an approximate formation temperature of $10^{5.0}$K. As proposed by \cite{Peter2000}, the core and wing components are indicative of small-scale transition region loops and large-scale coronal loops, respectively. 

\section{Conclusions}
\label{sec_conclu}
In this paper, we conduct a meticulous analysis of the spectral line characteristics of the \ion{Si}{IV} 1394 and 1403 \AA\ lines on the loop structure of a bifurcated eruption event. We employ multi-Gaussian profiles to fit the \ion{Si}{IV} 1394 \AA\ line at different times and positions on the loop, and calculate the integrated ratios $R$ as well as the intensity ratios along the wavelength points $r(\Delta\lambda)$ of the resonance lines. 

In the initial stage of the bifurcated eruption, the \ion{Si}{IV} line profiles of the loop exhibit obvious redshifts or blueshifts. From the double Gaussian fitting, it can be observed that these red (blue) velocities are provided by the dominant redshifted (blueshifted) components in the loop. In addition, the noticeable enhancement at both the red and blue wings can be observed before the eruption.

Around the moment of the bifurcated eruption, the emission of the \ion{Si}{IV} line shows a significant enhancement. At the northern positions (near loop L${_1}$), the \ion{Si}{IV} line profiles mostly show blueshifts, with a speed of approximately 10 to 20 \kms. On the other hand, at the southern side (near loop L${_2}$), the \ion{Si}{IV} line profiles exhibit noticeable red asymmetries, and obvious redshift components can be fitted with multiple Gaussian functions.

In the late stage of the bifurcated eruption, a distinct redshift component lasting several tens of seconds is observed near the origin position of the bifurcation loop (L${_1}$). Excluding the redshift velocity, the line profiles of the \ion{Si}{IV} line are blueshifted.

Moreover, it is observed that the majority of the integrated intensity ratio of \ion{Si}{IV} lines, denoted as $R$, exceeds 2. Furthermore, the ratio, $r(\Delta\lambda)$, which is measured along the wavelength points, exhibits noticeable variations across the wavelength spectrum. Despite these fluctuations, the ratio $r(\Delta\lambda)$ consistently maintains a value above 2. In addition to these observations, disparities in the Doppler velocities of \ion{Si}{IV} at 1394 Å and 1403 Å have been identified at certain positions.

In conclusion, we identify the ratio $r(\Delta\lambda)$ and $R$ as a significant metric for evaluating the resonance scattering and opacity of the \ion{Si}{IV} line during small AR eruption events.
The line profiles within these events exhibit complex structures, rendering the integrated ratio $R$ insufficient in fully encapsulating the opacity of the \ion{Si}{IV} line at varying wavelength. The opacity of the line core and line wings often diverges due to the presence of different velocity components within the line. The optically thin assumption of TR emission lines, which we usually consider, does not always hold during eruption events. And the opacity of the line center and line wings often exhibit significant differences. Thus the formation height of those lines may not correspond exactly to the position where its emissivity reaches its maximum.
When diagnosing the spectra of small AR events, it is crucial to consider not only the impact of opacity but also the influence of resonance scattering. The wavelength-dependent intensity ratio, denoted as $r(\Delta\lambda)$, proves to be a useful indicator for assessing the effect of resonance scattering. 
This conclusion necessitates further substantiation through additional observational and simulation evidence.

\begin{acknowledgements}We would like to sincerely thank the anonymous referee for his or her valuable feedback and constructive comments, which significantly improved the quality of this work. We would also like to thank the NVST, IRIS, and SDO teams for high-cadence data support. This work is sponsored by the Strategic Priority Research Program of the Chinese Academy of Sciences, Grant No. XDB0560000, the National Science Foundation of China( NSFC) under the numbers 12325303, 11973084, 11803085, 12003064, U1831210, 11803002, Yunnan Key Laboratory of Solar Physics and Space Science under the number 202205AG070009, Yunnan Provincial Science and Technology Department under the number 202305AH340002, and Yunling Scholar Project of Yunnan province.\end{acknowledgements}

\begin{appendix}
	\section{Additional figures of \ion{Si}{IV} line profiles}
\begin{figure*}[hb]\centering
		\includegraphics[width=2\columnwidth]{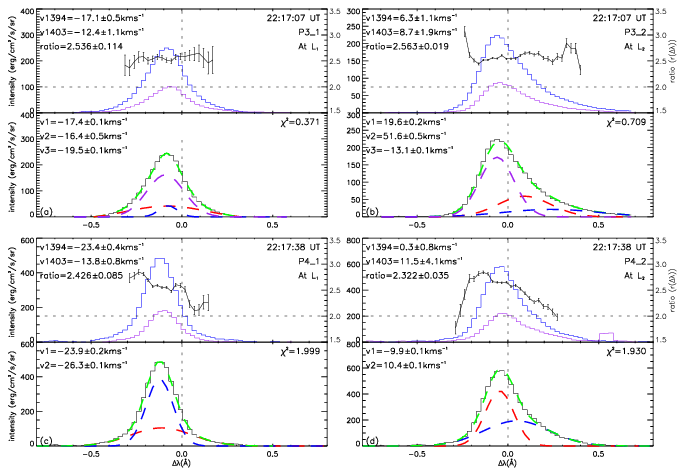}
		\caption{Same as Figure \ref{fig5}, but for positions during the eruption (P3\_1 - P4\_2).}
		\label{fig6}
	\end{figure*}

\begin{figure*}[hb]\centering
		\includegraphics[width=2\columnwidth]{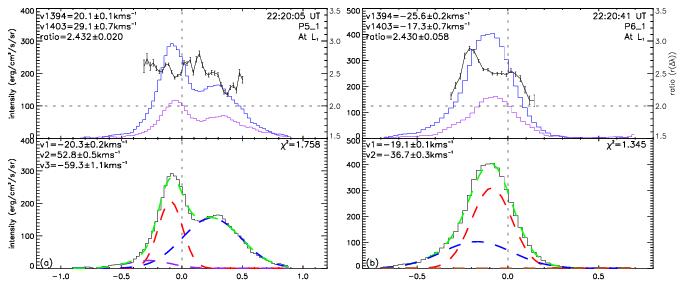}
		\caption{Same as Figure \ref{fig5}, but for positions after the eruption (P5\_1 - P6\_1).}
		\label{fig7}
	\end{figure*}

	Here, we present the profiles of the \ion{Si}{IV} resonance lines at loops during the eruption and in the later stage of the eruption. The lower panels in Figures \ref{fig6}{(a)}-{(d)} and \ref{fig7} show the observed \ion{Si}{IV} 1394 \AA\ line (histogram) and fitted line profiles (red, blue and purple) at each positions. The green dashed lines indicate the combined Gaussian fitting.
	\FloatBarrier
	
	\end{appendix}

\begin{thebibliography}{46}
\bibitem[Babu et al.(2024)]{Babu2024} Babu, B.~S., Kayshap, P., Tripathi, S.~C., et al.\ 2024, \mnras, 528, 2474. doi:10.1093/mnras/stae166

\bibitem[Bahauddin et al.(2021)]{Bahau2021} Bahauddin, S.~M., Bradshaw, S.~J., \& Winebarger, A.~R.\ 2021, Nature Astronomy, 5, 237. doi:10.1038/s41550-020-01263-2

\bibitem[Brannon et al.(2015)]{Brannon2015} Brannon, S.~R., Longcope, D.~W., \& Qiu, J.\ 2015, \apj, 810, 4. doi:10.1088/0004-637X/810/1/4

\bibitem[Chen et al.(2019)]{Chen2019a} Chen, Y., Tian, H., Huang, Z., et al.\ 2019, \apj, 873, 79. doi:10.3847/1538-4357/ab0417

\bibitem[Chen et al.(2019)]{Chen2019b} Chen, Y., Tian, H., Peter, H., et al.\ 2019, \apjl, 875, L30. doi:10.3847/2041-8213/ab18a4

\bibitem[Chitta et al.(2021)]{Chitta2021} Chitta, L.~P., Peter, H., \& Young, P.~R.\ 2021, \aap, 647, A159. doi:10.1051/0004-6361/202039969

\bibitem[Cook et al.(1995)]{Cook1995} Cook, J.~W., Keenan, F.~P., Dufton, P.~L., et al.\ 1995, \apj, 444, 936. doi:10.1086/175664

\bibitem[Dere et al.(1997)]{Dere1997} Dere, K.~P., Landi, E., Mason, H.~E., et al.\ 1997, \aaps, 125, 149. doi:10.1051/aas:1997368

\bibitem[De Pontieu et al.(2014)]{Depont2014} De Pontieu, B., Title, A.~M., Lemen, J.~R., et al.\ 2014, \solphys, 289, 2733. doi:10.1007/s11207-014-0485-y

\bibitem[Doschek(1984)]{Doschek1984} Doschek, G.~A.\ 1984, \apj, 279, 446. doi:10.1086/161907

\bibitem[Dud{\'\i}k et al.(2014)]{Dudik2014} Dud{\'\i}k, J., Del Zanna, G., Dzif{\v{c}}{\'a}kov{\'a}, E., et al.\ 2014, \apjl, 780, L12. doi:10.1088/2041-8205/780/1/L12

\bibitem[Gontikakis et al.(2013)]{Gontokakis2013} Gontikakis, C., Winebarger, A.~R., \& Patsourakos, S.\ 2013, \aap, 550, A16. doi:10.1051/0004-6361/200913423

\bibitem[Gontikakis \& Vial(2016)]{Gontokakis2016} Gontikakis, C. \& Vial, J.-C.\ 2016, \aap, 590, A86. doi:10.1051/0004-6361/201628109

\bibitem[Gontikakis \& Vial(2018)]{Gontokakis2018} Gontikakis, C. \& Vial, J.-C.\ 2018, \aap, 619, A64. doi:10.1051/0004-6361/201732563

\bibitem[Gupta \& Tripathi(2015)]{Gupta2015} Gupta, G.~R. \& Tripathi, D.\ 2015, \apj, 809, 82. doi:10.1088/0004-637X/809/1/82

\bibitem[Huang et al.(2014)]{Huang2014} Huang, Z., Madjarska, M.~S., Xia, L., et al.\ 2014, \apj, 797, 88. doi:10.1088/0004-637X/797/2/88

\bibitem[Huang et al.(2017)]{Huang2017} Huang, Z., Madjarska, M.~S., Scullion, E.~M., et al.\ 2017, \mnras, 464, 1753. doi:10.1093/mnras/stw2469

\bibitem[Kerr et al.(2019)]{Kerr2019} Kerr, G.~S., Carlsson, M., Allred, J.~C., et al.\ 2019, \apj, 871, 23. doi:10.3847/1538-4357/aaf46e


\bibitem[Landi et al.(2012)]{Landi2012} Landi, E., Del Zanna, G., Young, P.~R., et al.\ 2012, \apj, 744, 99. doi:10.1088/0004-637X/744/2/99

\bibitem[Lemen et al.(2012)]{Lemen2012} Lemen, J.~R., Title, A.~M., Akin, D.~J., et al.\ 2012, \solphys, 275, 17. doi:10.1007/s11207-011-9776-8

\bibitem[Maniak et al.(1993)]{Maniak1993} Maniak, S.~T., Tr{\"a}bert, E., \& Curtis, L.~J.\ 1993, Physics Letters A, 173, 407. doi:10.1016/0375-9601(93)90260-7

\bibitem[Mathioudakis et al.(1999)]{Math1999} Mathioudakis, M., McKenny, J., Keenan, F.~P., et al.\ 1999, \aap, 351, L23

\bibitem[Mulay \& Fletcher(2021)]{Mulay2021} Mulay, S.~M. \& Fletcher, L.\ 2021, \mnras, 504, 2842. doi:10.1093/mnras/stab367

\bibitem[M{\"u}ller et al.(2003)]{Muller2003} M{\"u}ller, D.~A.~N., Hansteen, V.~H., \& Peter, H.\ 2003, \aap, 411, 605. doi:10.1051/0004-6361:20031328

\bibitem[Nelson et al.(2017)]{Nelson2017} Nelson, C.~J., Freij, N., Reid, A., et al.\ 2017, \apj, 845, 16. doi:10.3847/1538-4357/aa7e7a

\bibitem[Pesnell et al.(2012)]{Pesnell2012} Pesnell, W.~D., Thompson, B.~J., \& Chamberlin, P.~C.\ 2012, \solphys, 275, 3. doi:10.1007/s11207-011-9841-3

\bibitem[Peter(2000)]{Peter2000} Peter, H.\ 2000, \aap, 360, 761

\bibitem[Peter et al.(2014)]{Peter2014} Peter, H., Tian, H., Curdt, W., et al.\ 2014, Science, 346, 1255726. doi:10.1126/science.1255726

\bibitem[Polito et al.(2020)]{Polito2020} Polito, V., De Pontieu, B., Testa, P., et al.\ 2020, \apj, 903, 68. doi:10.3847/1538-4357/abba1d

\bibitem[Raouafi et al.(1999)]{Raouafi1999} Raouafi, N.-E., Lemaire, P., \& Sahal-Br{\'e}chot, S.\ 1999, \aap, 345, 999

\bibitem[Rose et al.(2008)]{Rose2008} Rose, S.~J., Matranga, M., Mathioudakis, M., et al.\ 2008, \aap, 483, 887. doi:10.1051/0004-6361:20079040

\bibitem[Scherrer et al.(2012)]{Scherrer2012} Scherrer, P.~H., Schou, J., Bush, R.~I., et al.\ 2012, \solphys, 275, 207. doi:10.1007/s11207-011-9834-2

\bibitem[Tavabi et al.(2015)]{Tavabi2015} Tavabi, E., Koutchmy, S., \& Golub, L.\ 2015, \solphys, 290, 2871. doi:10.1007/s11207-015-0771-3

\bibitem[Teriaca et al.(2004)]{Teriaca2004} Teriaca, L., Banerjee, D., Falchi, A., et al.\ 2004, \aap, 427, 1065. doi:10.1051/0004-6361:20040503

\bibitem[Testa et al.(2014)]{Testa2014} Testa, P., De Pontieu, B., Allred, J., et al.\ 2014, Science, 346, 1255724. doi:10.1126/science.1255724

\bibitem[Tian et al.(2014)]{Tian2014} Tian, H., Li, G., Reeves, K.~K., et al.\ 2014, \apjl, 797, L14. doi:10.1088/2041-8205/797/2/L14

\bibitem[Tian et al.(2016)]{Tian2016} Tian, H., Xu, Z., He, J., et al.\ 2016, \apj, 824, 96. doi:10.3847/0004-637X/824/2/96

\bibitem[Tripathi et al.(2020)]{Tripathi2020} Tripathi, D., Nived, V.~N., Isobe, H., et al.\ 2020, \apj, 894, 128. doi:10.3847/1538-4357/ab8558

\bibitem[Vissers et al.(2015)]{Vissers2015} Vissers, G.~J.~M., Rouppe van der Voort, L.~H.~M., Rutten, R.~J., et al.\ 2015, \apj, 812, 11. doi:10.1088/0004-637X/812/1/11

\bibitem[Wang et al.(2023)]{Wang2023} Wang, L.~F., Li, Y., Li, Q., et al.\ 2023, \apjs, 268, 62. doi:10.3847/1538-4365/acf127

\bibitem[W{\"u}lser et al.(2018)]{Wulser2018} W{\"u}lser, J.-P., Jaeggli, S., De Pontieu, B., et al.\ 2018, \solphys, 293, 149. doi:10.1007/s11207-018-1364-8

\bibitem[Yan et al.(2015)]{Yan2015} Yan, L., Peter, H., He, J., et al.\ 2015, \apj, 811, 48. doi:10.1088/0004-637X/811/1/48

\bibitem[Young et al.(2015)]{Young2015} Young, P.~R., Tian, H., \& Jaeggli, S.\ 2015, \apj, 799, 218. doi:10.1088/0004-637X/799/2/218

\bibitem[Yu et al.(2023)]{Yu2023} Yu, H.~C., Hong, J., \& Ding, M.~D.\ 2023, \aap, 675, A171. doi:10.1051/0004-6361/202345931

\bibitem[Zhou et al.(2017)]{Zhou2017} Zhou, G.~P., Zhang, J., Wang, J.~X., et al.\ 2017, \apjl, 851, L1. doi:10.3847/2041-8213/aa9c40

\bibitem[Zhou et al.(2022)]{Zhou2022} Zhou, Y.-A., Hong, J., Li, Y., et al.\ 2022, \apj, 926, 223. doi:10.3847/1538-4357/ac497e

\end{thebibliography}
\end{document}